\documentclass[prc,showpacs,preprintnumbers,unsortedaddress,amsmath,amssymb,floatfix]{revtex4}
\input epsf.sty
\usepackage{graphicx}
\usepackage{bm}

\newcommand{\be}{\begin{equation}}
\newcommand{\ee}{\end{equation}}
\newcommand{\bea}{\begin{eqnarray}}
\newcommand{\eea}{\end{eqnarray}}

\begin{document}

\title{Properties of single-particle states in a fully self-consistent \\ particle-vibration coupling approach}

\author{Li-Gang Cao$^{1,2,3,4}$, G. Col$\grave{\text{o}}$$^{5,6}$, H. Sagawa$^{7,8}$, P. F. Bortignon$^{5,6}$}

\affiliation{${}^1$Institute of Modern Physics, Chinese Academy of
Science, Lanzhou 730000, China}

\address{${}^2$ State Key Laboratory of Theoretical Physics, Institute of Theoretical Physics, Chinese Academy of Sciences, Beijing 100190, China}

\address{${}^3$ Kavli Institute for Theoretical Physics China, CAS, Beijing 100190, China}

\affiliation{${}^4$ Center of Theoretical Nuclear Physics, National
Laboratory of Heavy Ion Accelerator of Lanzhou, Lanzhou 730000, P.R.
China}

\affiliation{${}^5$Dipartimento di Fisica,
Universit$\grave{\text{a}}$ degli Studi di Milano, via Celoria 16,
20133 Milano, Italy}

\affiliation{${}^6$Istituto Nazionale di Fisica Nucleare (INFN),
Sez. di Milano, via Celoria 16, 20133 Milano, Italy}

\affiliation{${}^7$Center for Mathematics and Physics, University of
Aizu, Aizu-Wakamatsu, Fukushima 965-8580, Japan}

\affiliation{${}^8$RIKEN, Nishina Center, Wako, 351-0198, Japan}

\begin{abstract}
The properties of single-particle states in the magic nuclei
$^{40}$Ca and $^{208}$Pb, in particular the energies, spectroscopic
factors and the effective mass, have been studied in a fully
self-consistent particle-vibration coupling (PVC) approach within
the framework of Skyrme energy density functional theory. All
selected phonons are obtained by the Random
Phase Approximation  and the same Skyrme interaction is also
used in the PVC vertex. 
We focus on the effect of the non-central two-body
spin-orbit and tensor interactions on the single-particle properties.   
It has been found that the contributions
of those terms are important to improve the results of
$^{208}$Pb. The calculated single-particle energies and
spectroscopic factors are compared to available experimental
data. The single-particle level density around the
Fermi surface is significantly increased due to the effect of PVC.
\end{abstract}

\pacs{21.10.Pc, 21.30.Fe, 21.60.Jz, 21.10.Jx}

\maketitle

\section{Introduction}
Since the 1970s, the self-consistent mean field (SCMF) approaches have
achieved a great success in describing various properties of
finite nuclei in their ground state, such as binding energies,
root-mean-square radii, and deformations \cite{Bender}.
The SCMF approaches have been extended to
describe the excited states, such as multipole giant resonances, and
rotational bands of finite nuclei. In those approaches, one starts in
general from an effective nucleon-nucleon interaction, such as
a Skyrme or Gogny interaction or a relativistic Lagrangian, and the
parameters of the effective interaction are fitted to the properties
of nuclear matter and some selected data of finite nuclei. The total binding
energy of a nucleus is expressed using as the integral of the energy density
that is, in turn, a function of the one-body densities; these are extracted from
the single-particle wave functions that are, with their corresponding energies,
obtained from the self-consistent solution of the
Schr$\ddot{\text{o}}$dinger or Dirac  equations. In such calculations,
the single-particle level density and the spectroscopic factors 
differ from the experimental findings mainly because of the
following reason. In the
mean field theory, the basic assumption is 
that particles move independently in the {\em static} average potential produced
by the surrounding nucleons. Of course, this assumption is an
ideal one. In practice, nucleons can make collisions with other
nucleons or couple to the collective vibrations of the whole system. This
is related to the concept of the so-called ``dynamical effects'' beyond
the mean field approximation. To consider the fluctuations of the mean field potential,
one must go beyond the mean field scheme, which means that
the average potential is no more static or energy-independent, and is instead
energy-dependent.

For finite nuclei, the fluctuations of the average
potential are usually described by an effective theory denoted as
``particle-vibration coupling'' (PVC) \cite{Boh75} model. It has been shown
that the particle-vibration coupling affects strongly the energies of
single-particle states around the Fermi surface and increases the
single-particle level density \cite{Ring73,Ham76,Bor77,Mah85}. In the
earlier times, the PVC calculations have been lacking
any self-consistency since the Woods-Saxon potential was usually
adopted to calculate the single-particle basis and the interactions
at the PVC vertex have been chosen with a large degree of
arbitrariness  \cite{Ring73,Ham76,Bor77,Mah85,Ber68,Wam82,Som83,Per80}. In few cases, the
Skyrme interaction has been adopted both for the single-particle potential
and PVC vertices, but using only the velocity-independent terms
in the PVC vertices \cite{Ber80}. Recently, microscopic
self-consistent PVC calculations have been performed within either the framework of
the Skyrme energy density functionals \cite{Colo10,Miz12,Bre12,Ogata} or the framework of
relativistic (i.e., covariant) functionals \cite{Lit06,Lit11}.

So far, even when the central terms of the Skyrme force have been consistently
included in the PVC vertex, the non-central terms such as
the two-body spin-orbit term or the tensor terms have been dropped in the calculations
\cite{Colo10}. At the same time, recently much attention
has been devoted to the tensor terms added to the Skyrme force with the goal to explain
e.g. the evolution of the single-particle levels
in exotic nuclei [based on Hartree-Fock (HF) or
Hartree-Fock-Bogoliubov (HFB) calculations]
\cite{Bro06,Colo07,Bri07,Les07,Gra07,Zal08,Ots05,Wang11,Dong11,tbp}.
Moreover, within HF plus self-consistent Random Phase
Approximation (RPA), some of us have investigated the effect of
the tensor force on the multipole response of finite nuclei
\cite{Cao11,Bai11}. The response function of uniform matter,
and the occurence of possible instabilities, has been the
subject of another recent study \cite{Pas121}. In the present work,
we shall study the effect of the non-central terms of
the Skyrme interactions at the PVC vertex on the single-particle
properties of finite nuclei. We will discuss the sensitivity of the
energy shifts associated with the single particle states, of
the effective mass, and of the spectroscopic factors, when the tensor
interaction and the spin-orbit interaction are included in the
PVC vertex. 
The calculations are performed
for the double magic nuclei $^{40}$Ca and $^{208}$Pb. The Skyrme
interactions adopted here are SLy5 \cite{Cha98} and
T44 \cite{Tij07}. For the case of SLy5, the terms associated with the
tensor force are simply added on top of the central force
as in Ref. \cite{Colo07},
whereas, in the case of T44, the
tensor parameters are fitted on the same footing as the other Skyrme
parameters. The ground states and the various excited states
of the nuclei $^{40}$Ca and $^{208}$Pb are calculated on the basis of
the fully self-consistent HF+RPA framework as in Ref. \cite{Colo12}. The
coupling of the particles to the vibrations is derived from the same
Skyrme force in a consistent way.

This paper is organized as follows. In Sec. II we will briefly
report the main features of our Skyrme HF plus RPA and PVC models,
as well as the definitions of other quantities which will be
discussed later. The results are displayed, analyzed and compared with
available experimental data in Sec. III. Section IV is devoted to
the summary and perspectives for future work(s).

\section{Method}
In this Section, we will briefly report the theoretical method adopted in
our calculations. More detailed information about the Skyrme HF
plus RPA calculations can be found in Ref. \cite{Colo12}.
First, we start by solving the Skyrme HF equations in the coordinate
space: the radial mesh is 0.1 (0.15) fm for $^{40}$Ca ($^{208}$Pb),
and the maximum value of the radial coordinate is set to be 15 (24)
fm for $^{40}$Ca ($^{208}$Pb), respectively. In order to calculate
unoccupied states at positive energy, the continuum has been
discretized by adopting box boundary conditions.  
In this way, we obtain
the energies as well as the wave functions for particle (p) and hole
(h) states, which are the input for RPA calculations. We solve the
RPA equations in the  matrix formulation; all the hole states
are considered when we build the particle-hole (p-h) configurations,
while for the particle states we choose the lowest six (eight)
unoccupied states for each value of $l$ and $j$ in the case, respectively, of
$^{40}$Ca ($^{208}$Pb). It has to be noted that, for $^{40}$Ca, RPA
produces instabilities if we include more than six shells when the
tensor force is considered. For $^{40}$Ca ($^{208}$Pb) we have
considered natural parity phonons with multipolarity $L$ ranging from 0 to 4
(from 0 to 5). For each  multipole response we have checked
that the RPA value of the
energy-weighted sum rule exhausts almost 100$\%$ of the analytic value calculated
from the double
commutator. 

After we obtain the RPA phonons, in our present PVC calculations
only those having energy smaller than 30 MeV and fraction of the
total isoscalar or isovector strength larger than 5$\%$ have been
considered for the coupling with single-particle states. In Table I
we present the properties of the low-lying states of $^{40}$Ca
and $^{208}$Pb, which give important contributions to the PVC
results (the available experimental data are also shown in Table I).
The results are obtained by using the SLy5 and T44 parameter sets
with and without considering the tensor force.
We can see that the tensor force affects in a substantial way both the
energies and the reduced transition probabilities of the low-lying
states of $^{40}$Ca and $^{208}$Pb.

The energy of the single particle (s.p.) state $i$ can be obtained by means
of second-order perturbation theory. We use such approach in the
present work. The
dressed single-particle energy $\varepsilon_i$ is expressed as
\begin{equation}
\varepsilon_i=\varepsilon_i^{(0)}+\Delta\varepsilon_i,
\end{equation}
where $\varepsilon_i^{(0)}$ is the single-particle energy given by
mean field calculations and $\Delta\varepsilon_i$ is the energy shift
calculated 
from 
the self-energy, that is,
\begin{equation}
\Delta\varepsilon_i=\Sigma_i(\omega=\varepsilon_i^{(0)}).
\end{equation}
The self-energy $\Sigma_i$ has the
following expression,
\begin{widetext}
\begin{equation}
\Sigma_i(\omega)=\frac{1}{2j_i+1}\left(\sum_{nL,p>F}\frac{|\langle i
\|V\|p,nL \rangle|^2}{\omega-\varepsilon_p^{(0)}-\omega_{nL}+i\eta} +
\sum_{nL,h<F}\frac{|\langle i \|V\|h,nL
\rangle|^2}{\omega-\varepsilon_h^{(0)}+\omega_{nL}-i\eta} \right),
\label{eq:self}
\end{equation}
\end{widetext} where $\varepsilon_p^{(0)}$ $(\varepsilon_h^{(0)})$ is the HF
single-particle (hole) energy,
and $\omega_{nL}$ is the energy of
phonon. The (small) imaginary part $\eta$ is set to be 0.05 MeV in our
calculations. The numerators contain the squared modulus of a
reduced matrix element called PVC vertex, which is expressed as
\begin{widetext}\begin{widetext}
\begin{table}
\caption{Energies and reduced transition probabilities of the
low-lying states in $^{40}$Ca and $^{208}$Pb obtained by HF
+ RPA with SLy5 and T44 parameter sets. The values in
parenthesis are the results obtained without the contribution
of the tensor force. The experimental data are from Ref. \cite{NNDC}.}
\begin{ruledtabular}
\begin{tabular}{cccccccc}

       &   & \multicolumn{4}{c}{Theory}               &\multicolumn{2}{c}{Exp.} \\
       &   &\multicolumn{2}{c}{SLy5} & \multicolumn{2}{c}{T44} &  &  \\
       &   &Energy &B(EL,0 $\rightarrow$ L) &Energy &B(EL,0 $\rightarrow$ L) & Energy &B(EL,0 $\rightarrow$ L)  \\
 &$J^\pi$   & [MeV]   & [e$^2$ fm$^{2L}$]   & [MeV]   & [e$^2$ fm$^{2L}$]     &  [MeV]   & [e$^2$ fm$^{2L}$]   \\

\hline

$^{40}$Ca & $3^-$   & 3.225(3.822)    & 0.884(1.285) $\times$ 10$^4$ & 1.366(1.508)    & 0.852(1.280) $\times$ 10$^4$ & 3.74  & 1.18 $\times$ 10$^4$\\

\hline

$^{208}$Pb & $2^+$  & 5.155(4.934)    & 3.065(2.858) $\times$ 10$^3$ & 4.549(5.105)    & 2.478(2.785) $\times$ 10$^3$ & 4.09  & 4.09 $\times$ 10$^3$\\


& $3^-$        & 3.585(3.671)   & 4.928(6.374) $\times$ 10$^5$ & 3.337(3.629)    & 5.739(5.523) $\times$ 10$^5$ & 2.61   & 6.21 $\times$ 10$^5$  \\
& $4^+$        & 5.760(5.417)   & 1.395(1.256) $\times$ 10$^7$ & 4.655(5.684)    & 0.782(1.382) $\times$ 10$^7$ & 4.32   & 1.29 $\times$ 10$^7$  \\
 & $5^-$       & 4.022(4.560)   & 2.881(4.898) $\times$ 10$^8$ & 3.977(4.092)    & 3.796(2.443) $\times$ 10$^8$ & 3.19   & 4.62 $\times$ 10$^8$  \\
         &     & 4.507(5.589)   & 0.748(1.642) $\times$ 10$^8$ & 4.532(5.021)    & 0.345(1.929) $\times$ 10$^8$ & 3.71   & 3.30 $\times$ 10$^8$   \\

\end{tabular}
\end{ruledtabular}
\end{table}
\end{widetext}\end{widetext}

\begin{widetext}
\begin{equation}
\langle i \|V\|j,nL \rangle=\sqrt{2L+1}\sum_{ph}X_{ph}^{nL}V_L(ihjp)
+(-)^{L+j_h-j_p}Y_{ph}^{nL}V_L(ipjh), \label{eq:PVCv}
\end{equation}
where $V_L$ is the particle-hole coupled matrix element,
\begin{equation}
V_L(ihjp)=\sum_{{\rm all}\ m}(-)^{j_j-m_j+j_h-m_h}\langle
j_im_ij_j-m_j|LM\rangle \langle j_pm_pj_h-m_h|LM\rangle \langle
j_im_i,j_hm_h|V|j_jm_j,j_pm_p\rangle. \label{eq:phma}
\end{equation}
\end{widetext}Details of the derivation of Eq. (3) can be found in
Ref. \cite{Colo10}.

The PVC effects are included in the energy-dependent self-energy
$\Sigma$.
In a uniform system, or in a finite system treated with the
local density approximation, the
single-particle energy 
can be written in a quite general fashion as
\begin{equation}
\varepsilon(k)=\frac{\hbar^2k^2}{2m}+\Sigma(k,\varepsilon(k)).
\label{eq:pair1}
\end{equation}
Here the self-energy $\Sigma$ includes both the HF potential
and the dynamical contributions from PVC (or, eventually,
further) correlations;
we have emphasized that such self-energy is
a function of the momentum
$k$ and energy $\varepsilon$. We can define an effective mass
$m^*$ through the relation:

\begin{equation}
\frac{m^*}{m}=\frac{\hbar^2k}{m}\left(\frac{d\varepsilon}{dk}\right)^{-1}.
\end{equation}

The momentum dependence of $\Sigma$ gives rise to
a non-locality, or $k$-mass
$\widetilde{m}$ which is related to $\Sigma$ by

\begin{equation}
\frac{\widetilde{m}}{m}=\left(1+\frac{m}{\hbar^2k}\frac{\partial
\Sigma}{\partial k}\right)^{-1}. \label{eq:pair3}
\end{equation}

The energy dependence of $\Sigma$ leads, instead, to a so-called
$E$-mass or $\omega$-mass, $\overline{m}$, defined by

\begin{equation}
\frac{\overline{m}}{m}=\left(1-\frac{\partial \Sigma}{\partial
\varepsilon}\right). \label{eq:pair4}
\end{equation}

Thus, the effective mass $m^*$ can be expressed in term of
$\widetilde{m}$ and $\overline{m}$,
\begin{equation}
\frac{m^*}{m}=\left(\frac{\widetilde{m}}{m}\right)_{\alpha}\times
\left(\frac{\overline{m}}{m}\right)_{\alpha}.
\label{eq:pair5}
\end{equation}
Since we deal in this work with finite nuclei, we have
stressed that these quantities are state-dependent by labelling them
with the quantum numbers
$\alpha$ 
of the HF single-particle state.
\begin{widetext}\begin{widetext}\begin{table}
\caption{The energies of the neutron single-particle states around
the Fermi surface in $^{40}$Ca calculated in various approximations.
The spectroscopic factors obtained in the full calculation are also shown in
this Table. The results are obtained by using SLy5 and T44 parameter
sets. The experimental data are taken from Refs. \cite{Arx07,Oros96}.}
\begin{ruledtabular}
\begin{tabular}{cccccccccccc}

& &  HF                 & \multicolumn{2}{c}{PVC}               &\multicolumn{2}{c}{PVC}&\multicolumn{2}{c}{PVC} & &\multicolumn{2}{c}{Spectroscopic}\\
& &                     & \multicolumn{2}{c}{central}            & \multicolumn{2}{c}{central+S.O.}& \multicolumn{2}{c}{full} & & \multicolumn{2}{c}{factors}\\
& &$\varepsilon^{(0)}$  &$\Delta\varepsilon_i$ &$\varepsilon_i$ &  $\Delta\varepsilon_i$ &$\varepsilon_i$&  $\Delta\varepsilon_i$ &$\varepsilon_i$ &  $\varepsilon_i^{exp}$ &$S_i^{th}$ &  $S_{i}^{exp}$ \\

\hline
SLy5 &$1f_{5/2}$  &  -1.26     & -1.36 & -2.62  & -1.07 & -2.33  & -2.11 & -3.37  &  -1.56 & 0.849   &0.95\\
     &$2p_{1/2}$  &  -3.11     & -1.95 & -5.06  & -1.54 & -4.65  & -2.04 & -5.15  &  -4.20 & 0.778   & 0.70\\
     &$2p_{3/2}$  &  -5.28     & -1.88 & -7.15  & -2.44 & -7.72  & -2.98 & -8.26  &  -5.84 & 0.823   & 0.91\\
     &$1f_{7/2}$  &  -9.69     & -0.83 & -10.52 & -1.30 & -10.99 & -1.56 & -11.26 &  -8.36 & 0.893   & 0.77\\

&&&&&&&&&&&\\
     &$1d_{3/2}$   &  -15.17    & -0.62 & -15.79 & -0.54 & -15.71 & -1.67 & -16.85 & -15.64 & 0.886  &0.94\\
     &$2s_{1/2}$   &  -17.26    & -1.13 & -18.39 & -1.51 & -18.77 & -2.12 & -19.38 & -18.19 & 0.845  &0.82 \\
     &$1d_{5/2}$   &  -22.10    & -0.31 & -22.41 & -0.65 & -22.75 & -1.07 & -23.17 & -20.39  & 0.923  &0.90\\
\hline

T44 &$1f_{5/2}$   & -0.21    & -2.00 & -2.21  & -2.59 & -2.80  & -2.67 & -2.88   &  -1.56 & 0.696   &0.95\\
    &$2p_{1/2}$   & -2.79    & -2.68 & -5.47  & -3.43 & -6.22  & -4.15 & -6.94   &  -4.20 & 0.773   & 0.70\\
    &$2p_{3/2}$   & -5.59    & -2.78 & -8.38  & -3.97 & -9.56  & -4.25 & -9.84   &  -5.84 & 0.676   & 0.91\\
    &$1f_{7/2}$   & -10.59   & -1.10 & -11.69 & -1.66 & -12.25 & -1.89 & -12.47  &  -8.36 & 0.815   & 0.77\\

&&&&&&&&&&&\\
    &$1d_{3/2}$   & -13.99   & -1.16 & -15.15 & -2.92 & -16.91 & -3.32 & -17.31  & -15.64 & 0.737  &0.94\\
    &$2s_{1/2}$   & -17.18   & -1.51 & -18.69 & -3.85 & -21.03 & -4.14 & -21.32  & -18.19 & 0.746  &0.82 \\
    &$1d_{5/2}$   & -22.59   & -0.49 & -23.08 & -0.78 & -23.37 & -0.74 & -23.34  & -22.39 & 0.772  &0.90\\
\end{tabular}
\end{ruledtabular}
\end{table}
\end{widetext}\end{widetext}

In particular, for a HF state, the $k$-mass ($\widetilde{m}/m)_{\alpha}$
can be written as
\begin{equation}
\left(\frac{\widetilde{m}}{m}\right)_{\alpha}= \int
|\varphi_{\alpha}(r)|^2\frac{\widetilde{m}(r)}{m}d^3r,
\label{eq:pair6}
\end{equation}
where $\widetilde{m}(r)$ is the effective mass 
associated with the given Skyrme set (which is
density-dependent and, therefore, radial-dependent because
of the nuclear density profile) while
$\varphi_{\alpha}$
is the Skyrme HF wave function. 

From the standard many-body theory, the energy-dependent self-energy
enters the Dyson equation for the single-particle Green's function
$G$, namely
\begin{equation}
(\varepsilon-\varepsilon_{\alpha}^{(0)}-\Sigma_{\alpha}(\varepsilon))
G_{\alpha}(\varepsilon)=1.
\label{eq:pair7}
\end{equation}
We work here in the so-called diagonal approximation, in which one
neglects the non-diagonal matrix elements $\Sigma_{\alpha\beta}$ on
the HF basis \cite{Ring73} and we label $\Sigma_{\alpha\alpha}$
simply by $\Sigma_\alpha$.
The poles of the Green's function $G_{\alpha}(\varepsilon)$
correspond to the zeros of
\begin{equation}
f(\varepsilon)=\varepsilon-\varepsilon_{\alpha}^{(0)}-
\Sigma_{\alpha}(\varepsilon),
\label{eq:pair8}
\end{equation}
and for each value of $\alpha$ there are several poles
$\varepsilon_{\alpha}^\lambda$ characterized by the index $\lambda$;
in other words,
because of the coupling to the collective vibrations the
single-particle state $\alpha$ becomes fragmented. In the vicinity of
a given
pole $\varepsilon_{\alpha}^\lambda$ the Green's function can be
represented (leaving aside a small ``background'' part) as
\begin{equation}
G_{\alpha}^{\lambda}(\varepsilon)=\frac{S_{\alpha}^{\lambda}}{\varepsilon-\varepsilon_{\alpha}},
\label{eq:pair9}
\end{equation}
where the residues at these poles correspond to the usual definition
of spectroscopic factors $S_{\alpha}^{\lambda}$, which is given by
\begin{equation}
S_{\alpha}^{\lambda}=\left( 1-\frac{\partial \Sigma_{\alpha}}{\partial
\varepsilon}\right)^{-1}_{\varepsilon=\varepsilon_{\alpha}^{\lambda}}.
\label{eq:pair10}
\end{equation}
The above Eqs. (12-15) are quite general.
In the current paper we stick, as already said, to perturbation theory
and the self-energy is calculated as in Eq. (3). Accordingly,
the spectroscopic factors of the above Eq. (15) are also
calculated only for the renormalized HF states, that is,
$\varepsilon_\alpha^\lambda$ is restricted to be
$\varepsilon_i^{(0)}$ of Eq. (1).
Such spectroscopic factors are displayed in Table II.
From the definition
of $S_{\alpha}^{\lambda}$, we can deduce the value of the
energy-dependent effective mass $(\overline{m}/m)_{\alpha}$
by making the inverse of the spectroscopic factor $S_{\alpha}^{\lambda}$.

\section{Results and Discussion}
In this Section we shall present our results for two nuclei:
$^{40}$Ca and $^{208}$Pb. The effective Skyrme interactions SLy5 and
T44 are used in our calculations. We will stress, in our discussion, the effect of
the non-central part of the Skyrme interaction (such as the
spin-orbit and tensor terms) on the single-particle energies
deduced from the PVC calculations.

\subsection{Results for $^{40}$Ca}

\begin{figure}[hbt]
\includegraphics[width=0.45\textwidth]{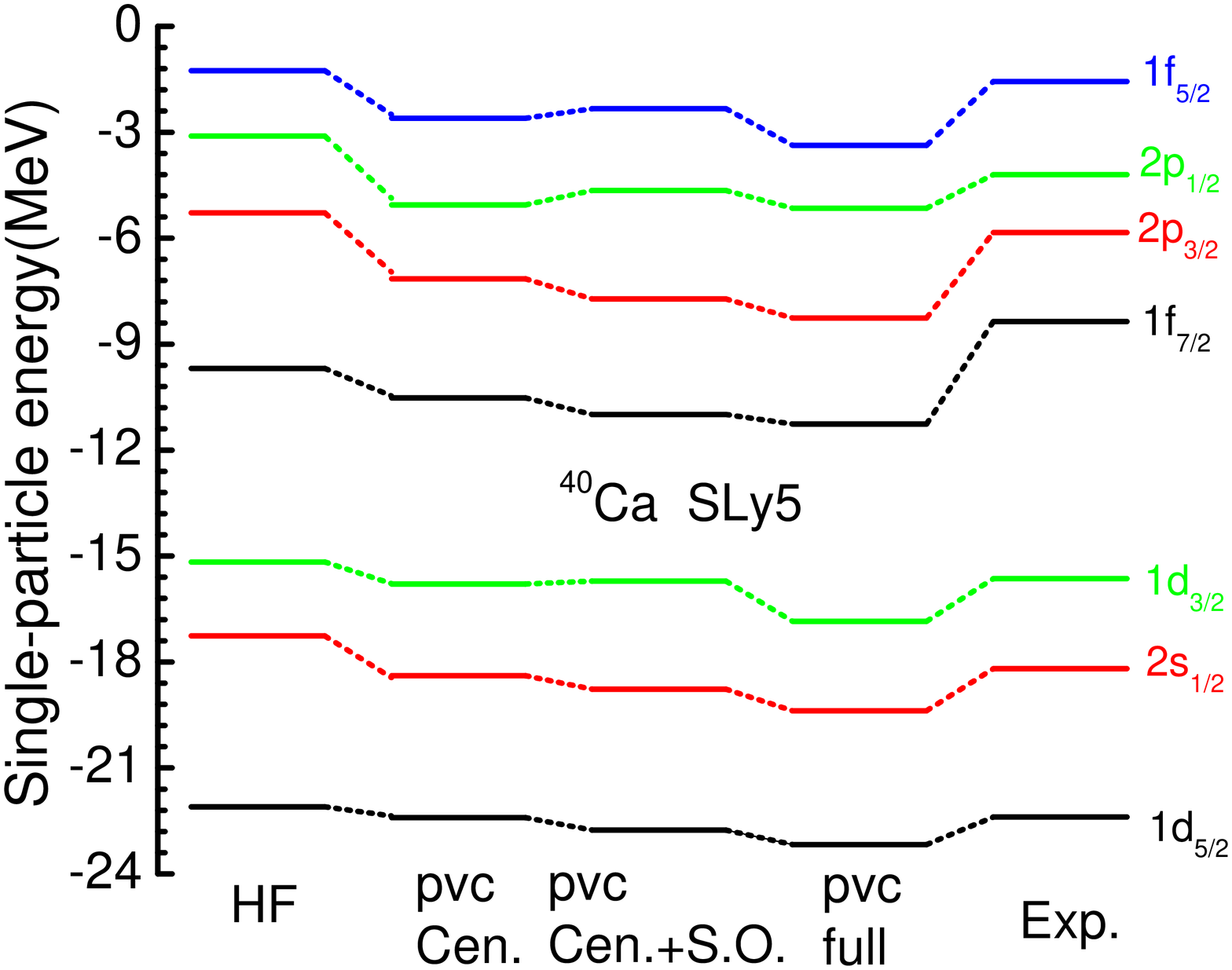}
\vglue
-2.cm\includegraphics[width=0.45\textwidth]{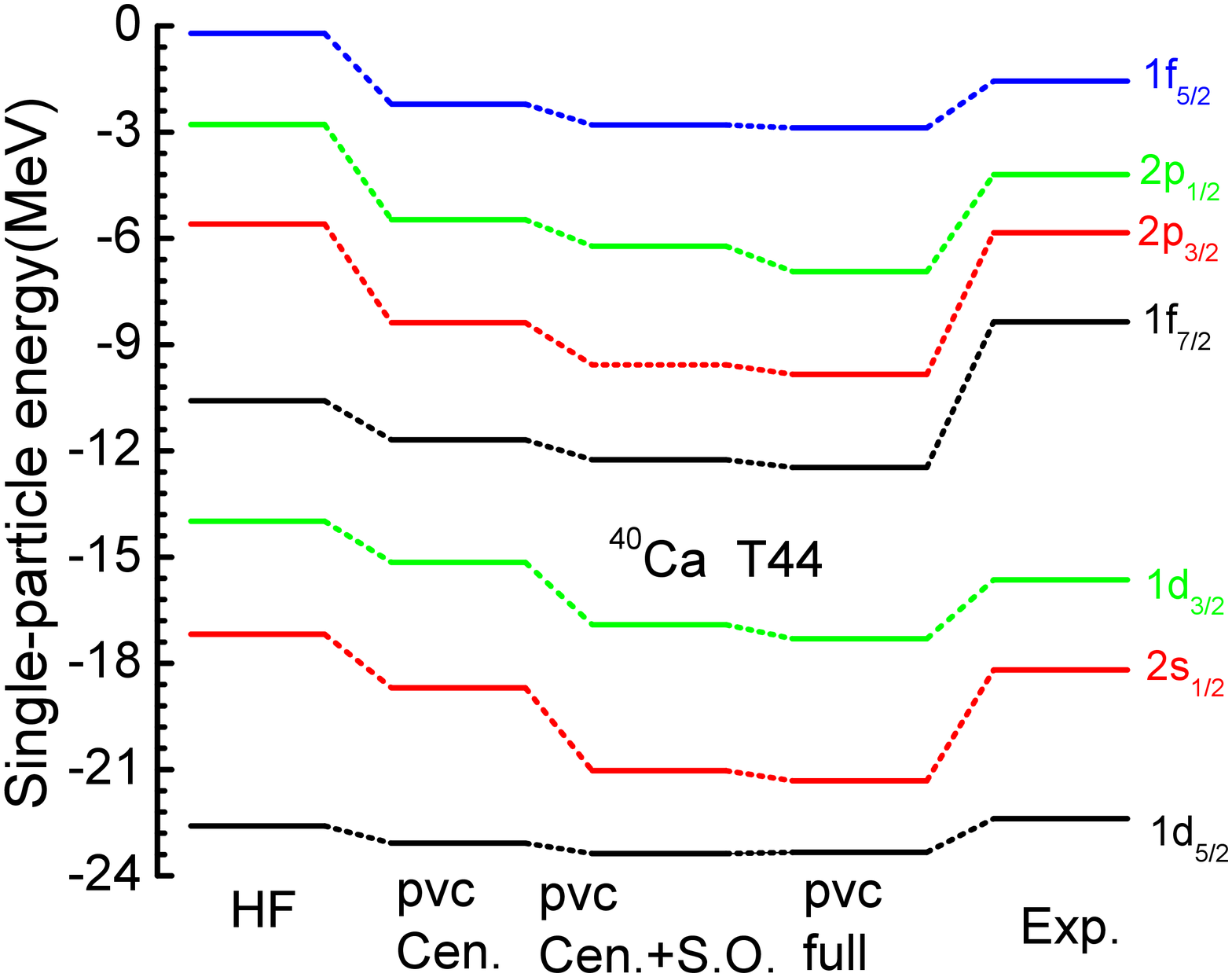}
\vglue -2.5cm \caption{(Color online) Neutron states in $^{40}$Ca
calculated with the parameter set SLy5 (upper panel) and T44 (lower
panel). The various columns, from left to right, correspond to the HF
calculation and to HF+PVC with only the central interaction in the vertex,
to HF+PVC with central and spin-orbit interaction in the
vertex, and to HF+PVC with central plus spin-orbit and tensor
interactions in the vertex. The last column displays the experimental data.}
\label{Fig.1}
\end{figure}

In Fig. 1 and Table II we show the results for the energies $\varepsilon_i$ of
neutron single-particle levels around the Fermi surface in $^{40}$Ca,
calculated in various approximation. The symbols are the same as in Eqs. (1) and (15).
The results (denoted by
$\varepsilon^{(0)}$) in the third column are obtained within the HF
mean field approximation by including the contribution of tensor
interaction, although the tensor interaction gives almost no
contribution to the single-particle energies in this case
since $^{40}$Ca is a $\vec l\cdot\vec s$-saturated nucleus (although we remind
that it affects the
energies and transition probabilities of the low-lying vibrations in
$^{40}$Ca \cite{Cao11}).
The results in
the fifth, seventh and ninth columns in Table II correspond to 
the PVC calculations 
with the central
Skyrme interaction, the central plus spin-orbit interaction and the central
plus spin-orbit as well as tensor interactions in the PVC vertex,
respectively. The values in the columns labelled by $\Delta\varepsilon_i$
are the difference between the PVC results and the original HF values.
The results are also compared with the available
experimental data.

The same information for the single-particle energies is shown in Fig. 1.
From Table II and Fig. 1, we can see that the single-particle energies, both below and
above the Fermi energy, become more negative when the
calculation includes the PVC effects. This qualitative outcome has
been already found and explained in Ref. \cite{Colo10}.
For the PVC results
obtained with only the central terms of the Skyrme force in the
vertex, the maximum energy shift is -1.95 MeV (-2.78 MeV) for
the $2p_{1/2}$ ($2p_{3/2}$) state using the SLy5 (T44) interaction.  
The spin-orbit interaction shows a repulsive effect on the energies of various
giant resonances in light nuclei,
and an attractive effect in heavy nuclei when included as residual interaction
in the RPA calculations. In the PVC
calculations, from Table II we see that it gives a repulsive
contribution to the energy shift for some states and an attractive
contribution for some other states in the case of the SLy5 force, while
if one moves to the T44 force, the spin-orbit interaction always gives an
attractive contribution for all single-particle states.
We will now discuss the contribution from tensor terms.
From Table II, we can see that the tensor force gives an attractive
contribution to the energy shift of all the single-particle levels, for both
the SLy5 and T44 Skyrme forces. We have also calculated the r.m.s.
deviation $\sigma$ between theoretical and experimental
single-particle states. The value of $\sigma$ is 1.026 (1.578), 1.330
(1.975), 1.566 (2.755), and 2.393 (3.010) in the case of HF, PVC
with central terms, PVC plus central and spin-orbit terms, and full PVC
calculation performed with
SLy5 (T44), respectively.
These results would go in the direction of calling for a re-fit of the Skyrme parameters.

In Table II, we also show the calculated
spectroscopic factors of single-particle states and the
corresponding experimental data. The results that we display are obtained by the full
calculation (all terms in the PVC vertex). For the SLy5 parameter set, the calculated
and measured values are more or less the same both for the particle
and hole states. For the T44 parameter set the calculated results are
systematically smaller than the experimental data for hole states.
For the particle states, the results do not show up a clear tendency.

\begin{table}
\caption{The calculated effective mass around the Fermi surface for
neutrons in $^{40}$Ca in various approximations. The results are
obtained by using the SLy5 and T44 parameter sets.}
\begin{ruledtabular}
\begin{tabular}{cccccccc}

& HF               & \multicolumn{2}{c}{PVC}               &\multicolumn{2}{c}{PVC}&\multicolumn{2}{c}{PVC} \\
&                     & \multicolumn{2}{c}{central}            & \multicolumn{2}{c}{central+S.O.}& \multicolumn{2}{c}{full} \\
        & $\widetilde{m}/m$ &$\overline{m}/m$&  $m^*/m$
        & $\widetilde{m}/m$ &$\overline{m}/m$&  $m^*/m$
\\

\hline
SLy5    & 0.852    & 1.091 & 0.931  & 1.107 & 0.944  & 1.170 & 0.999   \\
 T44    & 0.857    & 1.153 & 0.988  & 1.313 & 1.127  & 1.347 & 1.155   \\
\end{tabular}
\end{ruledtabular}
\end{table}

In Table III we show the effective $k$-mass, $E$-mass and the total
effective mass in $^{40}$Ca within various approximation.
These are obtained by averaging the effective masses associated
with the single-particle states that we have calculated (the averages
being, of course, done with the proper weights $2j_\alpha+1$).
The
effective $k$-mass is about 0.85 around the Fermi surface within the
pure Hartree-Fock mean field calculation for both the SLy5 and
T44 parameter sets. When one goes beyond the mean field calculation,
the mass operator is not only momentum-dependent but also
energy-dependent:
we can see that the calculated $E$--mass
is approximately in the range between 1.09 and 1.35 around the Fermi
surface.  
The effective mass, which is the product of $k$-mass and $E$-mass,
is $\approx$ 1.
We conclude
that the level density around the Fermi surface is enhanced when we
go beyond the mean field approximation using the PVC model.

\subsection{Results for $^{208}$Pb}

\begin{figure}[hbt]
\includegraphics[width=0.45\textwidth]{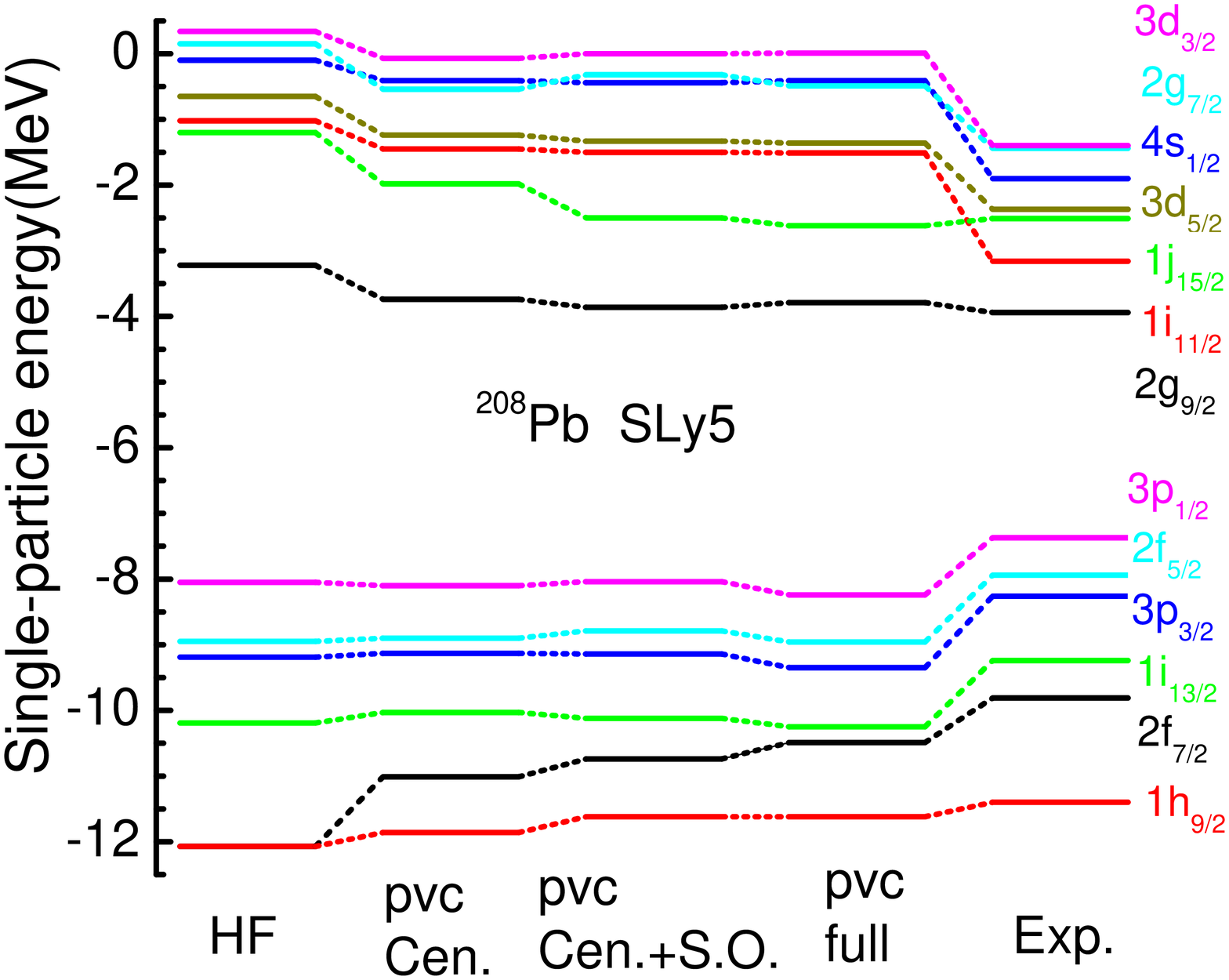}
\vglue
-2.cm\includegraphics[width=0.45\textwidth]{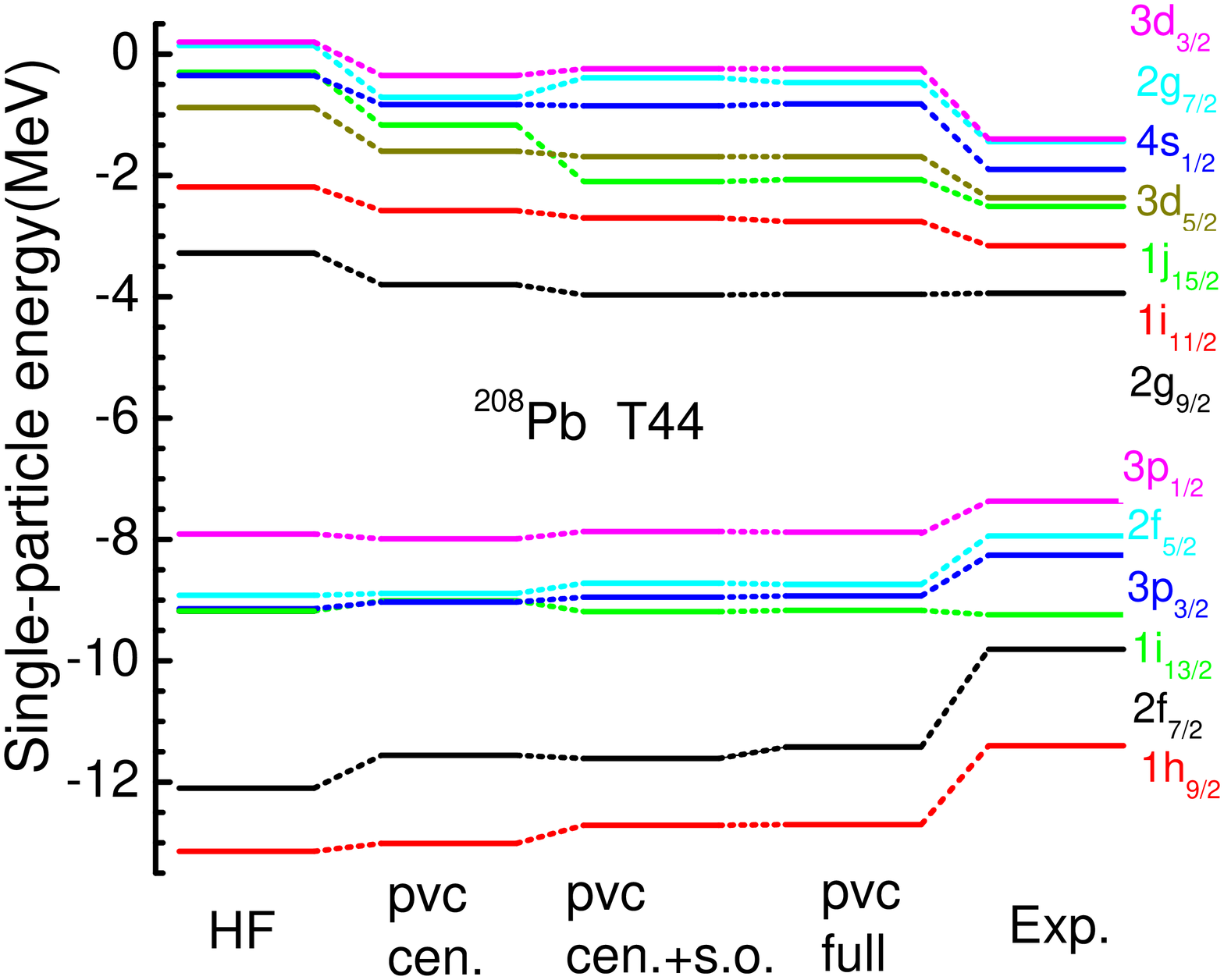}
\vglue -2.5cm \caption{(Color online) The same as Fig. 1 in the
case of the nucleus $^{208}$Pb.
}
\label{Fig.2}
\end{figure}

\begin{widetext}\begin{widetext}\begin{table}
\caption{The same as Table II
in the case of the nucleus $^{208}$Pb.
The experimental data are taken from Refs. \cite{Arx07,Oros96}.}
\begin{ruledtabular}
\begin{tabular}{cccccccccccc}

& & HF                 & \multicolumn{2}{c}{PVC}               &\multicolumn{2}{c}{PVC}&\multicolumn{2}{c}{PVC} & &\multicolumn{2}{c}{Spectroscopic}\\
& &                     & \multicolumn{2}{c}{central}            & \multicolumn{2}{c}{central+S.O.}& \multicolumn{2}{c}{full} & &  \multicolumn{2}{c}{factors}\\
& &$\varepsilon^{(0)}$  &$\Delta\varepsilon_i$ &$\varepsilon_i$ &  $\Delta\varepsilon_i$ &$\varepsilon_i$&  $\Delta\varepsilon_i$ &$\varepsilon_i$ &  $\varepsilon_i^{exp}$  &$S_i^{th}$ &  $S_{i}^{exp}$\\
 \hline
SLy5& $3d_{3/2}$   &   0.335    & -0.41 & -0.07  & -0.337& -0.002& -0.326&  0.009 &  -1.40& 0.911   &  1.09\\
    & $2g_{7/2}$   &   0.15     & -0.69 & -0.54  & -0.47 & -0.32 & -0.64 & -0.49  &  -1.44& 0.870   &  1.05\\
    & $4s_{1/2}$   &  -0.10     & -0.31 & -0.41  & -0.34 & -0.44 & -0.31 & -0.41  &  -1.90& 0.922   &  0.98\\
    & $3d_{5/2}$   &  -0.65     & -0.59 & -1.24  & -0.68 & -1.33 & -0.71 & -1.36  &  -2.37& 0.834   &  0.98\\
    & $1j_{15/2}$  &  -1.20     & -0.77 & -1.97  & -1.30 & -2.50 & -1.42 & -2.62  &  -2.51& 0.656   &  0.58\\
    & $1i_{11/2}$  &  -1.02     & -0.43 & -1.45  & -0.48 & -1.50 & -0.49 & -1.51  &  -3.16& 0.904   &  0.86\\
    & $2g_{9/2}$   &  -3.22     & -0.52 & -3.74  & -0.64 & -3.86 & -0.57 & -3.79  &  -3.94& 0.869   &  0.83\\

&&&&&&&&&&&\\
    & $3p_{1/2}$   &  -8.05     & -0.05 & -8.10  &  0.01 & -8.04 & -0.18 & -8.23  &  -7.37& 0.889   &  0.90\\
    & $2f_{5/2}$   &  -8.95     &  0.05 & -8.90  &  0.16 & -8.79 & -0.01 & -8.96  &  -7.94& 0.883   &  0.60\\
    & $3p_{3/2}$   &  -9.19     &  0.06 & -9.13  &  0.05 & -9.14 & -0.16 & -9.35  &  -8.26& 0.858   &  0.88\\
    & $1i_{13/2}$  &  -10.19    &  0.16 & -10.03 &  0.07 & -10.12& -0.06 & -10.25 & -9.24 & 0.908   &  0.91\\
    & $2f_{7/2}$   &  -12.07    &  1.07 & -11.00 &  1.33 & -10.74&  1.58 & -10.49 & -9.81 & 0.534   &  0.95\\
    & $1h_{9/2}$   &  -12.07    &  0.21 & -11.86 &  0.45 & -11.62&  0.45 & -11.62 & -11.40& 0.789   &  0.98\\
\hline
T44& $3d_{3/2}$   &  0.20     & -0.55 & -0.35  & -0.44 & -0.24 & -0.44 & -0.24 &  -1.40 & 0.895   &  1.09\\
   & $2g_{7/2}$   &  0.14     & -0.85 & -0.71  & -0.53 & -0.39 & -0.61 & -0.47 &  -1.44 & 0.832   &  1.05\\
   & $4s_{1/2}$   & -0.35     & -0.48 & -0.83  & -0.50 & -0.85 & -0.47 & -0.82 &  -1.90 & 0.896   &  0.98\\
   & $3d_{5/2}$   & -0.88     & -0.72 & -1.60  & -0.81 & -1.69 & -0.81 & -1.69 &  -2.37 & 0.855   &  0.98\\
   & $1j_{15/2}$  & -0.30     & -0.87 & -1.17  & -1.80 & -2.10 & -1.77 & -2.07 &  -2.51 & 0.583   &  0.58\\
   & $1i_{11/2}$  & -2.19     & -0.39 & -2.58  & -0.51 & -2.70 & -0.57 & -2.76 &  -3.16 & 0.884   &  0.86\\
   & $2g_{9/2}$   & -3.28     & -0.52 & -3.80  & -0.69 & -3.97 & -0.68 & -3.96 &  -3.94 & 0.877   &  0.83\\

&&&&&&&&&&&\\
   & $3p_{1/2}$   & -7.91     & -0.08 & -7.99  &  0.04 & -7.87 &  0.03 & -7.88 &  -7.37 & 0.905   &  0.90\\
   & $2f_{5/2}$   & -8.92     &  0.03 & -8.89  &  0.19 & -8.72 &  0.18 & -8.74 &  -7.94 & 0.888   &  0.60\\
   & $3p_{3/2}$   & -9.14     &  0.11 & -9.03  &  0.19 & -8.95 &  0.21 & -8.93 &  -8.26 & 0.844   &  0.88\\
   & $1i_{13/2}$  & -9.18     &  0.17 & -9.01  & -0.01 & -9.19 &  0.01 & -9.17 & -9.24  & 0.903   &  0.91\\
   & $2f_{7/2}$   & -12.10    &  0.54 & -11.56 &  0.49 & -11.61&  0.68 & -11.42& -9.81  & 0.580   &  0.95\\
   & $1h_{9/2}$   & -13.14    &  0.13 & -13.01 &  0.43 & -12.71&  0.44 & -12.70& -11.40 & 0.831   &  0.98\\
\end{tabular}
\end{ruledtabular}
\end{table}
\end{widetext}\end{widetext}

In Table IV we show the results for the energies $\varepsilon_i $ of
neutron single-particle levels around the Fermi surface in
$^{208}$Pb calculated within various approximation, exactly as in the
case of $^{40}$Ca that we have just discussed.
The results
given by the HF mean-field approximation, denoted by
$\varepsilon^{(0)}$ in the third column, are obtained by including the
contribution of the tensor interaction.  
There are finite
contributions of the tensor terms to the single-particle energies in the
ground state of $^{208}$Pb which is not a $\vec l\cdot\vec s$-saturated nucleus.
The results are compared with the available experimental data. The same
theoretical and experimental energies are displayed in
Fig. 2.

From Table IV and Fig. 2, we can see that the PVC
calculations give a small repulsive contribution to the energy for
most of the hole states below the Fermi surface. One noticeable exception is the $2f_{7/2}$
hole state which is shifted up in energy by about 1.60 MeV in the case of the SLy5
parameter set, and by 0.70 MeV in the case of the T44 parameter set. On the other hand, for
its spin-orbit partner state $2f_{5/2}$ the energy shift is rather
small. This goes against the prejudice that spin-orbit partner states
should be affected more or less in the same way by the PVC effects:
in this case, the special role of the coupling with the low-lying 3$^-$
breaks this ``rule of thumb''.
For the particle states, the energy shift $\Delta
\varepsilon_i$ is always negative but its magnitude depends on
the state chosen and on the approximation scheme.
In particular, the spin-orbit
and tensor terms of the force do not give a systematic effect: 
for some states they give a repulsive contribution to the energy shift
(with respect to the shift obtained by retaining only the central part
of the Skyrme force at the PVC vertex), whereas for some other states
they give attractive contributions. We have also calculated the
r.m.s. deviation $\sigma$ between theoretical and experimental
single-particle states. The values of $\sigma$ are 1.451, 1.030,
0.993, 1.097 for SLy5 and 1.421, 1.002, 0.907, 0.873 for T44 in the
case of HF, PVC with central terms, PVC with central plus spin-orbit terms, and full PVC
calculations, respectively.
In this case the inclusion of all terms in the PVC vertex produces an
improvement of the results, although small.

In Table IV we also show the calculated
spectroscopic factors of the single-particle states and the
corresponding experimental data. The calculations are performed within our
full PVC model. 
For the particle states, the agreement between the calculated and the
measured spectroscopic factors is generally
satisfactory. For the hole states, the largest disagreement between
theoretical and experimental data is found in the case of the $2f_{5/2}$ and
$2f_{7/2}$ states. For the $2f_{5/2}$ state, we can see that it is rather
fragmented from the experimental side, whereas the calculated fragmentation
is rather small. For its spin-orbit partner state
$2f_{7/2}$, the situation is opposite.

\begin{table}
\caption{The same as Table III in the case of the nucleus
$^{208}$Pb.
}
\begin{ruledtabular}
\begin{tabular}{ccccccccc}

&  HF                 & \multicolumn{2}{c}{PVC}               &\multicolumn{2}{c}{PVC}&\multicolumn{2}{c}{PVC} \\
&                      & \multicolumn{2}{c}{central}            & \multicolumn{2}{c}{central+S.O.}& \multicolumn{2}{c}{full} \\
        & $\widetilde{m}/m$ &$\overline{m}/m$&  $m^*/m$
        & $\widetilde{m}/m$ &$\overline{m}/m$&  $m^*/m$
\\
\hline
SLy5  & 0.839    & 1.156 & 0.968  & 1.198 & 1.002 & 1.229 & 1.028 \\
T44   & 0.841    & 1.157 & 0.973  & 1.200 & 1.009 & 1.235 & 1.038 \\
\end{tabular}
\end{ruledtabular}
\end{table}

In Table V, we show the effective $k$-mass, $E$-mass and total
effective mass in $^{208}$Pb.
As in the previous case of $^{40}$Ca, we have averaged the effective
masses of the states that we have considered (above and below the Fermi
surface).
The
effective $k$-mass is about 0.84 around the Fermi surface in the
pure HF calculation with the SLy5 and T44 parameter
sets. When one goes beyond the mean field calculation, 
we see that the calculated $E$-mass is approximately between
1.16 and 1.24 around the Fermi surface. The total effective mass, that is,
the product of the $k$-mass and $E$-mass, 
of the states around the Fermi surface is about
one (which is comparable to the empirical value). 

\section{Summary}
In this paper, the properties of the
single-particle states, in particular the energies, the
spectroscopic factors and the effective masses, in the magic nuclei
$^{40}$Ca and $^{208}$Pb, have been studied in a fully
self-consistent particle-vibration coupling (PVC) approach within the
framework of Skyrme energy density functional theory. All the
vibrations (phonons) are produced within a fully self-consistent Random
Phase Approximation (RPA) scheme.  
The SLy5 and T44 parameter sets are adopted; the tensor terms are
added to the central terms without any re-fit
in the case of the SLy5 parameter set.  

We have paid a specific attention to the effect produced on the single-particle
properties by the non-central part of the Skyrme
interaction. 
It has been found, in the case of the single-particle energies,
that the contributions to their energy shift induced by the tensor and spin-orbit
terms are smaller than those coming from the central Skyrme terms. In the case of the spin-orbit
terms, the contribution to the single-particle energy shift is
quite random for both $^{40}$Ca and $^{208}$Pb, namely it can be either
positive or negative. The contribution to the energy shifts stemming from the tensor force is
negative in $^{40}$Ca , 
while it has a random sign for
$^{208}$Pb. 

For $^{208}$Pb using the set T44, our results are improved with respect to the experimental findings by
the contributions of spin-orbit and tensor forces. The
calculated single-particle energies and spectroscopic factors
show an overall good agreement with data. This is reflected in the
enhancement of the single-particle
level density around the Fermi surface, due to the PVC correlations.
The effective mass becomes indeed close to one around the Fermi
energy, and this is consistent with the empirical information.

This work is a further step in the direction of improving mean-field
models when the need to be compared with the single-particle states and
their fragmentation. The role of higher order process, beyond our simple
perturbation theory approach, should be investigated. More importantly,
we should see if more significant improvements in the agreement between
theory and experiment can be obtained when the effective force is
re-fitted.

The coupling between single-particle and collective degrees of freedom
is also important if one wants to describe
of the optical potential that characterized finite nuclei when
projectile nucleons interact in e.g. a scattering process.
The particle-vibration coupling can provide an important contribution to
the imaginary part of the
optical potential. Work in this direction is also in
progress.

\section*{Acknowledgments}
L.G. Cao acknowledges the support of the National Science Foundation
of China under Grant Nos. 10875150 and 11175216, and the Project of
Knowledge Innovation Program of Chinese Academy of Sciences under
Grant No KJCX2-EW-N01. This work is partially supported by the
Japanese Ministry of Education, Culture, Sports, Science and
Technology by Grant-in-Aid for Scientific Research under the program
number (C(2))20540277. The support of the Italian Research Project
"Many-body theory of nuclear systems and implications on the physics
of neutron stars" (PRIN 2008) is also acknowledged.


\begin{thebibliography}{99}





\bibitem{Bender} M. Bender, P.-H. Heenen, P.-G. Reinhard, Rev. Mod. Phys.
\textbf{75} (2003) 121.

\bibitem{Boh75} A. Bohr and B. R. Mottelson, Nuclear structure. Vol.
II (W. A. Benjamin, New York, 1975).

\bibitem{Ring73} P. Ring, and E. Werner, Nucl. Phys. {\bf A211} (1973) 198.

\bibitem{Ham76} I. Hamamoto and P. Siemens, Nucl. Phys. {\bf A269} (1976) 199.

\bibitem{Bor77} P.F. Bortignon, R.A. Broglia, C.H. Dasso,
C. Mahaux, Phys. Lett. {\bf B140} (1984) 163.

\bibitem{Mah85} C. Mahaux,  P. F. Bortignon, R. A. Broglia, and C. H. Dasso,
Phys. Rep. {\bf 120} (1985) 1.

\bibitem{Ber68} G. F. Bertsch and T.T.S. Kuo, Nucl. Phys. {\bf A112} (1968) 204.

\bibitem{Wam82}  J. Wambach, V. K. Mishra and Li Chu-Hsia, Nucl. Phys. {\bf A380} (1982) 285.

\bibitem{Som83}  H. M. Sommermann, K. F. Ratcliff and T.T.S. Kuo, Nucl. Phys. {\bf A406} (1983) 109.

\bibitem{Per80}  R.P.J. Perazzo, S.L. Reich and H.M. Sofia, Nucl. Phys. {\bf A339} (1980) 23.

\bibitem{Ber80}  V. Bernard and N. V. Giai, Nucl. Phys. {\bf A348} (1980) 75.

\bibitem{Colo10} G. Col\`o,  H. Sagawa, and P. F. Bortignon, Phys. Rev. {\bf C82} (2010) 064307.

\bibitem{Miz12}  K. Mizuyama, G. Col\`o, and E. Vigezzi, Phys. Rev. {\bf C86} (2012) 034318.

\bibitem{Bre12}  M. Brenna, G. Col\`o, and P. F. Bortignon, Phys. Rev. {\bf C85} (2012) 014305.

\bibitem{Ogata} K. Mizuyama and K. Ogata, Phys. Rev. {\bf C86} (2012) 041603.

\bibitem{Lit06} E. Litvinova and P. Ring, Phys. Rev. {\bf C73} (2006) 044328.

\bibitem{Lit11} E. Litvinova and A.V. Afanasjev, Phys. Rev.
{\bf C84} (2011) 014305.

\bibitem{Bro06} B. A. Brown, T. Duguet, T. Otsuka, D. Abe,
and T. Suzuki, Phys. Rev. C {\bf 74} (2006) 061303(R).

\bibitem{Colo07} G. Col\`o, H. Sagawa, S. Fracasso and P.F.
Bortignon, Phys. Lett. B {\bf 646} (2007) 227 [see also: Phys. Lett.
B {\bf 668} (2008) 457].

\bibitem{Bri07} D. M. Brink and F. Stancu, Phys. Rev. C {\bf 75}
(2007) 064311.

\bibitem{Les07} T. Lesinski, M. Bender, K. Bennaceur, T. Duguet,
and J. Meyer, Phys. Rev. C {\bf 76} (2007) 014312.

\bibitem{Gra07} M. Grasso, Z. Ma, E. Khan, J. Margueron, and
N. Van Giai, Phys. Rev. C {\bf 76} (2007) 044319.

\bibitem{Zal08} M. Zalewski, J. Dobaczewski, W. Satula, and
T. R. Werner, Phys. Rev. C {\bf 77} (2008) 024316.

\bibitem{Ots05} T. Otsuka, T. Suzuki, R. Fujimoto, H. Grawe,
and Y. Akaishi, Phys. Rev. Lett. {\bf 95} (2005) 232502; T. Otsuka,
T. Matsuo, and D. Abe, Phys. Rev. Lett. {\bf 97} (2006) 162501.

\bibitem{Wang11} Y. Z. Wang, J. Z. Gu, X. Z. Zhang, et. al., Pyhs. Rev. C {\bf 84} (2011) 044333.

\bibitem{Dong11} J. M. Dong, W. Zuo, X. Z. Zhang, et. al., Pyhs. Rev. C {\bf 84} (2011) 014303.

\bibitem{tbp} H. Sagawa, G. Col\`o, Prog. Part. Nucl. Phys. (submitted).

\bibitem{Cao11}  L. G. Cao, H. Sagawa, and G. Col\`o, Phys. Rev. {\bf C83} (2011)
034324; L. G. Cao, G. Col\`o, and H. Sagawa,Phys. Rev. {\bf C81} (2010)
044302; L. G. Cao, G. Col\`o, and H. Sagawa, {\em et al.}, Phys. Rev.
{\bf C80} (2009) 064304.

\bibitem{Bai11}  C. L. Bai, H. Q. Zhang, H. Sagawa, {\em et al.}, Phys. Rev.
{\bf C83} (2011) 054316; C. L. Bai, H. Q. Zhang, H. Sagawa, {\em et al.}, Phys. Rev. Lett. {\bf
105} (2010) 072501.

\bibitem{Pas121}  A. Pastore, M. Martini, V. Buridon, D. Davesne, K. Bennaceur, and J. Meyer, Phys. Rev.
{\bf C86}, 044308 (2012); A. Pastore, D. Davesne, Y. Lallouet, M.
Martini, K. Bennaceur, and J. Meyer, Phys. Rev. {\bf C85}, 054317
(2012); D. Davesne, M. Martini, K. Bennaceur, and J.Meyer, Phys. Rev.
{\bf C80}, 024314 (2009); {\bf C84}, 059904(E) (2011).

\bibitem{Cha98} E. Chabanat, P. Bonche, P. Haensel, J. Meyer, and R. Schaeffer, Nucl. Phys. {\bf A635}, 231 (1998).

\bibitem{Tij07} T. Lesinski, M. Bender, K. Bennaceur, T. Duguet, J. Meyer,
Phys. Rev. {\bf C76}, 014312 (2007).

\bibitem{Colo12} G. Col\`o, L.G. Cao, N. Van Giai, L. Capelli,
Comp. Phys. Comm. {\bf 184}, 142 (2013).

\bibitem{NNDC} [http://www.nndc.bnl.gov].

\bibitem{Arx07} N. Schwierz, I. Wiedenh$\ddot{\text{o}}$ver, and A. Volya, arXiv:0709.3525.

\bibitem{Oros96} A. Oros, Ph.D. thesis, University of K$\ddot{\text{o}}$ln, 1996.

\end{thebibliography}
\end{document}